\documentclass[eqnum,showpacs,showkeys,prd]{revtex4}
\usepackage{amsmath}
\begin{document}

\title{Area and Entropy: A New Perspective}
\author{Jarmo M\"akel\"a}
\email{jarmo.makela@puv.fi}
\affiliation{Vaasa Polytechnic, Wolffintie 30 FIN-65200 Vaasa, Finland}
\begin{abstract}

We consider a spacelike two-plane originally at rest with respect to electromagnetic radiation in thermal equilibrium. We find that if the plane is moved with respect to the radiation, the plane shrinks such that the maximum amount of entropy carried by radiation through the plane is, in natural units, exactly one-half of the decrease in the area of the plane. This result suggests that the equivalence between area and entropy may not be limited in black holes, nor even in the spacetime horizons only, but the equivalence between horizon area and entropy may be just a special case of some general and simple, still undiscovered fundamental principle of nature.

\end{abstract}
\maketitle

One of the most important specific problems of modern theoretical physics is posed by the Bekenstein-Hawking entropy law. That law states that black hole has entropy $S$, which is proportional to its event horizon area $A$.\cite{eka} More precisely, in SI units:
\begin{equation}
S = \frac{1}{4}\frac{k_Bc^3}{\hbar G}A.
\end{equation}
During the past 10 years or so there have been numerous attempts, mostly based on string theory and loop quantum gravity, to give a microscopic, statistical derivation for this law.\cite{toka} It is commonly believed that the origin of Eq.(1) lies in the microscopic, quantum-mechanical degrees of freedom of spacetime, and that if we only had a perfect understanding of these microscopic degrees of freedom, we would be very close to the quantum theory of gravitation.

When facing a physical problem, it is often useful to look at that problem from different angles. When considering the Bekenstein-Hawking entropy law, an obvious question is: Is that law limited in black holes or, more generally, in spacetime horizons only, or is it possibly a special case of some general and simple principle of nature? After all, the Bekenstein-Hawking entropy law is a thermodynamical law, and the laws of thermodynamics are both general and simple. The question is therefore: Is there some simple, still undiscovered law of thermodynamics which would yield, among other things, the Bekenstein-Hawking entropy law?\cite{koo}

We begin our investigations with an observation that when a black hole is in thermal equilibrium with its surroundings, the radiation processes of the hole are reversible.\cite{nee} In that case the entropy of the hole is exactly converted to the entropy of the radiation, and between the entropy $dS$ carried by radiation out of the hole, and the change $dA$ in the horizon area of the hole there is, in the natural units, where $\hbar= c=G=k_B=1$, the relationship:
\begin{equation}
dS = -\frac{1}{4}\,dA.
\end{equation}
This equation gives us an idea to seek for a general relationship between the amount of entropy carried by radiation through a spacelike two-surface, and the change in the area of that two-surface. The key question is: In which way does the entropy of radiation interact with the geometry of spacetime?

      An assumption of thermal equilibrium between the hole and its surroundings is essential in the derivation of Eq.(2). Because of that we consider first a spacetime region filled with electromagnetic radiation in thermal equilbrium. In this spacetime region we have a spacelike two-plane with area $A$ at rest with respect to the radiation, and an observer at rest with respect to the two-plane. From the point of view of our observer the energy density of the radiation is
\begin {equation}
\rho = T_{\mu\nu}\xi^\mu\xi^\nu,
\end{equation}
and its pressure is
\begin{equation}
p = \frac{1}{3}\rho = T_{\mu\nu}\eta^\mu\eta^\nu.
\end{equation}
In these equations $\xi^\mu$ is the future pointing unit tangent vector of the observer's world line, and $\eta^\mu$ is a spacelike unit normal vector of the plane. The vector $\eta^\mu$ is assumed to be orthogonal to $\xi^\mu$, and because the observer is at rest with respect to the radiation, there is no net flow of energy nor momentum through the two-plane, and we have:
\begin{equation}
T_{\mu\nu}\xi^\mu\eta^\nu = 0.
\end{equation}

So far we have assumed an exact thermal equilibrium from the point of view of our observer. We now introduce a  very small perturbation for this equilibrium: We {\it move} the two-plane during a very short interval of time. When the plane moves with respect to the radiation with certain speed, radiation, and therefore entropy, flows through the plane. Before the plane has achieved a certain speed, however, the plane must have been {\it accelerated} first. If we accelerate the plane with a constant proper acceleration $\kappa$ during a very small proper time interval $\tau$ to the direction of the vector $-\eta^\mu$, the vectors $\xi^\mu$ and $\eta^\mu$ transform to the vectors $\xi'^\mu$ and $\eta'^\mu$ such that
\begin{subequations}
\begin{eqnarray}
\xi'^\mu &=& \xi^\mu + \kappa\tau\,\eta^\mu,\\
\eta'^\mu &=& \kappa\tau\,\xi^\mu + \eta^\mu,
\end{eqnarray}
\end{subequations}
and the heat $\delta Q$ carried through the two-plane to the direction of the vector $\eta^\mu$ during an infinitesimal proper time interval $d\tau$ is:\cite{vii}
\begin{equation}
\delta Q = AT_{\mu\nu}\xi'^\mu\eta'^\nu\,d\tau = \frac{4}{3}\kappa A\rho\tau\,d\tau,
\end{equation}
where we have used Eqs.(3)-(5). Because $\tau$ is assumed to be very small, we have neglected in Eqs.(6) and (7) the terms non-linear in $\tau$.

 The change in the area $A$ of the two-plane during the infinitesimal proper time interval $d\tau$ may be calculated by means of the Raychaudhuri equation\cite{kuu}
\begin{equation}
\frac{d\theta}{d\tau} = a^\mu_{;\mu} - \frac{1}{3}\theta^2 - \sigma_{\mu\nu}\sigma^{\mu\nu} + \omega_{\mu\nu}\omega^{\mu\nu} + R_{\mu\nu}u^\mu u^\nu.
\end{equation}
 In this equation $\theta$, $\sigma_{\mu\nu}$ and $\omega_{\mu\nu}$, respectively, are the expansion, shear and twist of a congruence of timelike curves of spacetime. $u^\mu$ is the future pointing unit tangent vector field of this congruence, and $a^\mu:=u^\alpha u^\mu_{;\alpha}$ is the proper acceleration vector field of the congruence. When we use the Raychaudhuri equation for the evaluation of the change in the area $A$ of the two-plane, we consider a congruence which consists of the world lines of the points of the plane. For that congruence $a^\mu a_\mu=constant:=\kappa^2$, and we may choose $\theta=\sigma_{\mu\nu}=\omega_{\mu\nu}=0$ and $u^\mu=\xi^\mu$, when $\tau=0$. So the Raychaudhuri equation takes the form:
\begin{equation}
\frac{d\theta}{d\tau} = a^\mu_{;\mu} + R_{\mu\nu}\xi^\mu\xi^\nu,
\end{equation}
when $\tau=0$.

       The geometrical meaning of the expansion $\theta$ is that if we consider a congruence of timelike curves which consists of the world lines of the points of a small three-dimensional region of spacetime, the change in the volume $V$ of that region during an inifinitesimal proper time interval $d\tau$ is:
\begin{equation}
dV = V\theta\,d\tau.
\end{equation}
Using Eq.(9) we therefore find that if we pick up a small spacelike region from the neighborhood of the spacelike two-plane under consideration, the change in the volume of that region during an infinitesimal proper time interval $d\tau$ is:
\begin{equation}
dV = Va^\mu_{;\mu}\tau\,d\tau - 8\pi V\rho\tau\,d\tau,
\end{equation}
where we have, again, neglected the terms non-linear in $\tau$. When obtaining Eq.(11) we have used Eq.(3), together with Einstein's field equation
\begin{equation}
R_{\mu\nu} = -8\pi(T_{\mu\nu} - \frac{1}{2}g_{\mu\nu}T^\alpha_\alpha),
\end{equation}
and the fact that $T^\alpha_\alpha = 0$ for electromagnetic radiation in thermal equilibrium. The first term on the right hand side of Eq.(11) is due to mere acceleration, and the contribution it gives to $dV$ is caused by the change in the spatial geometry in a direction orthogonal to the spacelike two-plane. In other words, mere acceleration does not cause changes in the area of the two-plane. Those changes may be read off from the remaining term on the right hand side of Eq.(11), which is due to the interaction between spacetime geometry and radiation. Since the radiation is in thermal equilibrium, and therefore homogeneous, we may assume that the spatial region under consideration contracts or expands in the same way in all directions. Moreover, because the area of a two-boundary of any three-dimensional region of space is proportional to the power 2/3 of the volume of that region, we may infer that the change in the area of the accelerating two-plane during the infinitesimal proper time interval $d\tau$ is:
\begin{equation}
dA = -\frac{2}{3}\,8\pi A\rho\tau\,d\tau = -\frac{16\pi}{3}A\rho\tau\,d\tau.
\end{equation}
Finally, using the first law of thermodynamics,
\begin{equation}
\delta Q = T\,dS,
\end{equation}
and comparing Eqs.(7) and (13) we find that the entropy $dS$ carried by radiation with temperature $T$ through the two-plane may be expressed as:
\begin{equation}
dS = -\frac{1}{T}\frac{\kappa}{4\pi}\,dA.
\end{equation}
In other words, the two-plane {\it shrinks} while radiation goes through the plane such that the entropy carried by radiation through the plane is proportional to the decrease in its area.

             So far our treatment has been purely classical. At this point quantum mechanics enters the stage. According to the so-called {\it Unruh effect} an acccelerating observer with proper acceleration $\kappa$ will observe a steady flow of thermal radiation with the {\it Unruh temperature}\cite{seite}
\begin{equation}
T_U := \frac{\kappa}{2\pi}
\end{equation}
even when the radiation fields are in vacuum  from the point of view of all inertial observers. The Unruh temperature $T_U$ may therefore be viewed as the minimum temperature of radiation an accelerating observer may measure. Substituting $T_U$ for $T$ in Eq.(15) we may hence obtain the maximum value for the amount of entropy carried by radiation in thermal equilibrium through a spacelike two-plane:
\begin{equation}
dS_{max} = -\frac{1}{2}\,dA.
\end{equation}
As one can see from Eq.(2), this is exactly {\it twice} the amount of entropy carried by radiation in thermal equilibrium out from a black hole.\cite{kasi}  

Eq.(17) is very remarkable. It tells what happens if a thermal equilibrium of radiation is slightly perturbed such that a two-plane originally at rest with respect to the radiation is moved a bit. According to Eq.(17) the plane shrinks, and the maximum amount of entropy carried by radiation through the two-plane during its movement is, in natural units, exactly one-half of the decrease in the area of the plane. This result clearly points to the possibility that the equivalence between area and entropy may not be limited in black holes, nor even in the horizons of spacetime only. Eq.(17) therefore provides a new perspective for the relationship between area and entropy. It may even suggest an existence of some new, still undiscovered principle of thermodynamics which would set, in certain conditions, an ultimate upper bound for the amount of entropy carried by radiation when it interacts with spacetime. It is possible that if such principle existed, then the Bekenstein-Hawking entropy law, among other things, might be just one of the consequences of that principle. At this stage  of research it is too early to express speculations about the precise form of that principle. Whatever the fundamental principle suggested by the similarities between the Eqs.(2) and (17) may be,  its search will certainly be of great fun!

\end{document}